*研究論文*

# Prospective Study for Semantic Inter-Media Fusion in Content-Based Medical Image Retrieval

Roxana TEODORESCU[*1,2], Daniel RACOCEANU[*3],
Wee-Kheng LEOW[*4], Vladimir CRETU[*1]

**Abstract**

　One important challenge in modern Content-Based Medical Image Retrieval (CBMIR) approaches is represented by the semantic gap, related to the complexity of the medical knowledge. Among the methods that are able to close this gap in CBMIR, the use of medical thesauri/ontologies has interesting perspectives due to the possibility of accessing on-line updated relevant web-services and to extract real-time medical semantic structured information. The CBMIR approach proposed in this paper uses the Unified Medical Language System's (UMLS) Metathesaurus to perform a semantic indexing and fusion of medical media. This fusion operates before the query processing (retrieval) and works at an UMLS-compliant conceptual indexing level. Our purpose is to study various techniques related to semantic data alignment, preprocessing, fusion, clustering and retrieval, by evaluating the various techniques and highlighting future research directions. The alignment and the preprocessing are based on partial text/image retrieval feedback and on the data structure. We analyze various probabilistic, fuzzy and evidence-based approaches for the fusion process and different similarity functions for the retrieval process. All the proposed methods are evaluated on the Cross Language Evaluation Forum's (CLEF) medical image retrieval benchmark, by focusing also on a more homogeneous component medical image database: the Pathology Education Instructional Resource (PEIR).

Med Imag Tech **26**(1): -, 2008

## 1. Introduction

　In the medical field, digital images are produced in huge quantities and used for direct diagnosis and therapy. Even though the introduction of DICOM[*5] medical image format standardization and PACS[*6] medical information storage and management systems represent important milestones in the medical field, much effort is needed to use these standards efficiently and effectively for diagnosis assistance, teaching and research.

　In the same way that PACS expands on the possibilities of a conventional hard-copy medical image storage system by providing capabilities of off-site viewing and reporting (distant education, telediagnosis) and by enabling practitioners at various physical locations to access the same information simultaneously (teleradiology), Content-Based Medical Image Retrieval (CBMIR) opens the gate to the next generation of medical procedures. For instance, CBMIR systems could provide advanced diagnosis assistance, setup semantic links between the related medical information for improving patient health care. Furthermore, in the near future, data-mining could be used for research applications, for the medical queries expansion and for all the potential Evidence-Based Medicine (EBM) and Image Based Reasoning (IBR) [1, 2] applications generated by the similarity-based image retrieval. Finally, decision support systems in radiology and computer-aided diagnostics for radiological practice need powerful data and meta-data management and retrieval [3, 4].

Even if the image-based queries will most likely not be able to replace text-based ones, they have the potential to be a very good complement to text-based search, based on their visual characteristics. Accordingly, we propose a general semantic CBMIR system using a web-available medical metathesaurus - UMLS (Unified Medical Language System[*7]), by treating the fusion between the conceptual indexes of the medical images and medical reports. This approach has the advantage of using an up-to-date medical metathesaurus (updates are published many times a year), taking into account many languages, a huge number of biomedical concepts and essential symbolic and statistic relations between them. This structured medical metathesaurus offers the opportunity of homogeneous fusion between UMLS-compliant concepts coming from different medical media (images, reports ...), but also automatic query expansion and rule extraction.

This paper is organized as follows: section 2 summarizes a brief state-of-the-art in CBMIR, focusing on actual challenges like the semantic gap and the use of complementary medical media. The following section is dedicated to the fusion methodology, including a comparative study of probabilistic, fuzzy and evidence-based theoretical techniques. A study of a post-fusion clustering that is able to decrease the retrieval processing time, with perspectives to create medically meaningful clusters, is introduced in the section 4. Similarity measures and techniques are studied in section 5 and the results obtained on the CLEF 2006 medical image retrieval benchmark are presented in section 6. The study ends with the main conclusions and future research directions in section 7.

## 2. Brief overview of content-based medical image retrieval systems

Content-based image retrieval (CBIR) is the application of computer vision to the image retrieval problem, i.e., the problem of searching for digital images in large databases. "Content-based" means that the search makes use of the contents of the images themselves, rather than relying on textual annotation or human-input metadata.

The visual features used for indexing and retrieval are classified in [5] into three classes:
－ primitive features that are low-level features such as color, shape and texture;
－ logical features that are medium-level features describing the image by a collection of objects and their spatial relationships;
－ abstract feature that are semantic and contextual features.

The obvious loss of information from image data to a representation by abstract features is called the semantic gap [6] and constitutes nowadays a major research topic in this field. In order to close this semantic gap to improve retrieval performances, specialized retrieval systems have been proposed in literature. Indeed, the more specialized the application is for a limited domain, the smaller the gap can be made using domain knowledge [7~9]. Nonetheless, the concepts for medical image retrieval are limited to a particular modality, organ, or diagnostic study and, hence, usually not directly transferable to other medical applications [10].

In [10] the authors propose a general content-based Image Retrieval in Medical Applications (IRMA). This system aims at developing and implementing high-level methods for CBMIR with applications in medical-diagnosis tasks on radiological image archive. Based on a general structure for semantic image analysis that results in 6 layers of information modeling, IRMA is implemented with distributed system architecture suitable for large databases. Until now, this system was only used for basic queries regarding the category of images (modality, orientation, body region, biological system).

Another important reference in CBMIR systems is the medGIFT project [11]. Textual information as well as structured information plays a much more important role in the medical field. Pre-treatment of images is performed.

---

[*7] UMLS - http://www.nlm.nih.gov/research/umls/

This includes the removal of the background to concentrate research on the main part of the image as well as segmentation of specific objects for retrieval.

Efficient and discriminative content-based indexing of medical imaging still remains a difficult challenge. Nowadays, classical visual parameters extraction methods have reached a limit, relating to the use of the medical knowledge. Integrating the medical domain knowledge into the indexing and retrieval algorithms will certainly constitute one of the main research topics of the CBMIR.

Our CMBIR approach uses Metathesaurus UMLS Knowledge Source. The medical report indexes, represented by corresponding UMLS concepts have associated frequency/inverse frequency parameters. According to the one-versus-all (OVA) approach, Support Vector Machine (SVM) classifiers have been used for the global images to obtain UMLS concepts based on a supervised learning set. The objective of our work is situated after the partial media indexing by focusing on a homogeneous way of fusing the semantic UMLS indexes of the medical images and medical reports, in order to enhance the performances of CBMIR system.

### 3. Inter-media fusion

Steinberg et al. [12] define data fusion as *the process of combining data to refine state estimates and predictions*, suggesting that data fusion contains information fusion and sensor fusion, the difference between the two being made by the process of correlation or estimation used.

Composed of a collection of medical reports and a collection of medical images, our database represents a *medical multimedia database*. In the fusion process, the starting point is given by the data obtained from the image and text indexing processes. Through the fusion process, we obtain a combination of concepts that we refer to as *medical case indexes*.

Certain characteristics are extracted from the medical image, whereas some specific elements can only be extracted from the text documents such as medical reports. By trying to extract the similar medical cases based only on medical images (MAP[*8] 9%) or on the medical report (MAP 17%) we loose part of the information. By fusing the common concepts from the medical image(s) corresponding to a certain medical report, we try to give more importance to those concepts, because they are found in both sides. In this case the information is not lost, but it is used for a better extraction and the specific concepts complete each other. The fusion presented in this article consists of five steps: Pre-processing, Fusion, Similarity computation, Content-Based Medical Image Retrieval part and Evaluation of results. All these steps are developed in the following subsections. As shown in **Fig. 1** the input data in this system from the image part represents the *concept identifiers* (from UMLS) and the trust degree generated by the SVMs. Each UMLS Concept Unique Identifier (CUI) extracted from the medical reports has an associated tf/idf (term frequency / inverse document frequency) value. The common concepts are then fused using one of

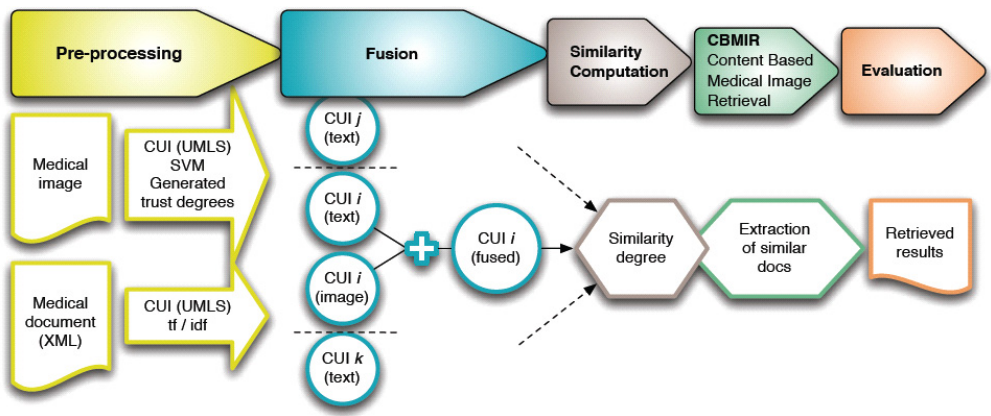

**Fig. 1** Fusion diagram flowchart.

---

[*8] The mean average precision-defined in section 3

the fusion operators. All the concepts, the *fused* ones and those that are not common, are used to find the documents that are similar with the ones in the query. Those documents are stored in the database as *dictionaries* that contain the concepts from the medical report and their attached image(s). These dictionaries are classified using the *clustering* technique presented in section 4. When we look for a specific document, the *similarity computation* searches for the documents that have the most concepts in common with the ones found in the query. The value given by the similarity computation part reveals the similar cases with the query. Using the identifiers from the dictionaries the actual images and the XML documents can be displayed and shown in the last step of the system.

**1）Preprocessing**

In the clinical practice, a medical case constitutes one or more medical reports and one or more associated medical images. In our approach, we consider decomposition into elementary medical cases c formed by one medical report and one associated medical image. The combination of the elementary cases can give a reconstruction of the original medical case.

The elementary medical case c thus includes indexing of the associated image and medical report:

$$\Lambda_{txt}^c = \begin{bmatrix} CUI_{i_1}, & \lambda_{txt_{i_1}}^c \\ ... & ... \\ CUI_{i_n}, & \lambda_{txt_{i_n}}^c \end{bmatrix}, \Lambda_{img}^c = \begin{bmatrix} CUI_{j_1}, & \lambda_{img_{j_1}}^c \\ ... & ... \\ CUI_{j_m}, & \lambda_{img_{j_m}}^c \end{bmatrix} \quad (1)$$

$$with: \lambda_{txt_{i_l}}^c = \mu_{txt_{i_l}}^c * \nu_{txt_{i_l}}^c * \omega_{txt_{i_l}}^c * \varphi_{txt_{i_l}}^c, l=1,...,n \quad (2)$$

$$\lambda_{img_{j_q}}^c = \mu_{img_{j_q}}^c * \nu_{img_{j_q}}^c * \omega_{img_{j_q}}^c * \varphi_{img_{j_q}}^c, q=1,...,m \quad (3)$$

where: *CUI* is the UMLS's Concept Unique Identifier, $\mu$ the fuzzy confidence degree, $\nu$ the relative frequency of the concept, $\omega$ the spatial localization fuzzy weight and $\varphi$ the data test feedback or relevance feedback.

The *spatial localization* $\omega$ corresponds, in the medical report indexing, to the importance of the section to which the concept belongs. For example, in the < *Diagnosis* > paragraph of the medical report, the physician included the most important keywords for describing the disease (the pathology) and the anatomic part. This tag will thus be more important than the < *Description* > tag, which is more neutral. For the medical image, $\omega$ is used to enhance a particular region of interest having a relevant meaning for a given modality such as anatomical part or a pathology.

The *feedback* $\varphi$ represents the confidence accorded to the extracted concepts. Indeed, for medical image indexing, some of the SVMs have a better performances for certain types of modality detection, naturally yielding a better associated feedback. There are two types of feedback coefficients: possibility measures, that take into account the results from a test dataset classification and relevance feedback coefficients. We adopt the first type in our application.

The *confidence degree* $\mu$ is a fuzzy result directly given by the SVM classifiers, according to the quality of the training dataset used for medical image indexing. For the concepts extracted from the text part, the value for this parameter is 1, since the existence or the absence of the concepts is not fuzzy.

For the text pre-processing, a text indexing software has been used to calculate the *local relative frequency* $\nu_{txt_i}^c$ of the concept occurrence in the given medical report. If a patch extraction method is used for the image, the local relative frequency $\nu_{img_i}^c$ will be computed using the relative weight of this concept versus all the patches of the image. In our case, as the classifiers act on the global images (we have not used patches), the value of the $\nu_{img_i}^c$ parameter is 1. Indeed, for each image, a global generalization approach has been used, corresponding to the definition of the SVM. For the text concepts, this parameter is computed as *tf/idf* (term frequency/inverse document frequency) or it can be computed as *drf* (document related frequency).

In order to better understand the pre-processing stage and to have a view of the granularity of the data, the table presented in **Fig. 2** contains some indexed images and the extracted SVM values correlated with the CUIs (in our database we have 32 possible image types, so 32 CUIs will always be present in the image indexing file). The CUIs correspond to UMLS concepts and each image modality has a CUI.

On the text part (**Fig. 3**), each medical concept extracted, corresponds to a UMLS CUI and has an attached tf/idf value. Globally, the set of medical reports, generate many more concepts extracted than from the image files (from the MIR[*9] database, the smallest, we had 56,000 CUIs). Each medical report (from the 50,000 cases of the CLEF Database) generates an average of about 50 UMLS CUIs. For each medical report, there can be one or more image(s) attached; a medical report along with its attached image(s) is called a "case". The medical cases from our database look like the example in **Fig. 4**. In this case, for the XML file the four images correspond to it. The Figure represents the indexed images and medical reports in the way they are used as input into our system.

2） **Data alignment**

After data preprocessing an alignment is necessary in order to balance the influence of each component media (text and image) in the final result. We propose the use of the medium feedback given by Interpolated Recall – Precision Averages ($rp_{txt}$ and $rp_{img}$) obtained on partial media (text respectively image) retrieval:

$$\Lambda'_{txt} = \Lambda_{txt}\alpha_{txt} \ \ and \ \ \Lambda'_{img} = \Lambda_{img}\alpha_{img} \tag{4}$$

where:

$$\frac{\alpha_{txt}}{\alpha_{img}} = \frac{avg_{img} \cdot rp_{txt}}{avg_{txt} \cdot rp_{img}} \tag{5}$$

The alignment method based on the partial media retrieval feedback aims at balancing the two datasets depending on their individual retrieval (recall and precision) performances. As far as we know, this idea is a new and a generic method that can considerably increase the quality of the retrieval. In section 6, we will introduce our results and conclusion about this important topic (see **Fig. 6**).

3） **Fusion Approach**

There are several fusion methods in literature, depending on the data that is provided and on the final purpose of the fusion. Different classification criteria have been proposed, from the point of view of the nature of the data and respectively from the data quality.

**Low, Intermediate and High Level**

— *Low level fusion - data fusion* takes several sources of data and combines them into a new data row, more informative or more synthetic than any of the original ones ［13］；
— *Intermediate level fusion - feature-level fusion* combines various features. It can be used to fuse the features or to find significant features among the ones extracted in a row of data. Methods for this type of fusion include *Principal Component Analysis* (*PCA*), *Multi-layer Perceptrons* (*MLP*), *etc*.

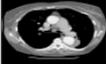

**Fig. 2** Example of indexed query images with some UMLS's Concept Unique Identifiers (CUI) extracted.

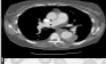

**Fig. 3** Example of indexed medical query text.

---

[*9] gamma.wustl.edu/home.html

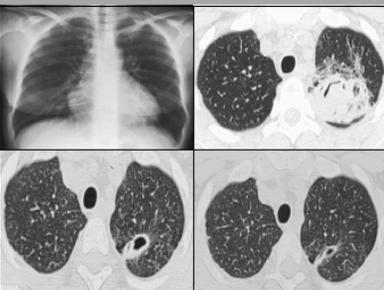

**Fig. 4** Example of a medical case and its indexed values.

― *High level - decision fusion* combines decisions coming from several experts. The experts can also return a confidence degree. These methods include *voting methods, statistical methods, fuzzy logic methods*.

**Probabilistic, Fuzzy and Evidence-Based Fusions**. These types of fusion are used for decision making when the information is imprecise, uncertain or incomplete. Data fusion imposes in general three major tasks: sensor registration/data gathering, data association and data combination.

Typical association techniques include nearest neighbor search (which selects the closest item), probabilistic data association (which selects a set of candidates with confidence weights) and multiple hypothesis (association decision making is postponed until enough evidences are collected).

The *Probabilistic*, *Fuzzy* and *Evidence Based* fusion techniques have their own operators, depending on the data provided and on their features, as well as on the purpose of the fusion.

In our approach we consider the *Context Independent Constant Behavior* (*CICB*) 〔14〕 operators. Since they are stable, we do not need the contextual information or external information.

If we refer to operators as disjunctive and conjunctive, then the conjunctive operators create a consensus between the data, giving more confidence to the source which has the smallest measure. This kind of operators searches for a simultaneous satisfaction of the criteria. The disjunctive operator increases the confidence in the source with the most certain measure, or the one with the biggest value. A compromise operator provides global measure, intermediating between the partial measures given by the sensors. Redundancy can be overcome by fuzzy operators.

Wald 〔15〕 suggests that Dempster-Shafer has the best performance in the decision fusion, the last level of fusion. The highest level of fusion has a decision as the result, whereas our level of fusion, the intermediate level, has feature fusion as a result. We use the operators represented in **Table 1**, where the x and y values represent the computed $\Lambda$ values from equation 4 for the text, respectively the image extracted concepts. The first two types of operators are included in the fuzzy algebra and the reason for choosing these operators is given in the third column. The mean operator and the symmetrical sums represent the probabilistic approach.

The Mean Average Precision (MAP) presented in **Table 1** represents the value of the relevance given to the documents retrieved by the system:

$$MAP = N^{-1} \sum_{i=1}^{N} q_i \qquad (6)$$

Table 1 Selected fusion operators.

| Operators type | Operators | Reason | MAP(CLEF) | MAP(PEIR) |
|---|---|---|---|---|
| T-conorms | max(x,y) | the smallest T-conorm | 25.54% | 27.17% |
|  | min(1,x+y) | saturation as soon as (x+y)>1 | 25.59% | 27.28% |
| T-norms | min(x,y) | the biggest T-norm | 23.66% | 25.65% |
|  | max(0,x+y-1) | not discriminant | 23.60% | 26.08% |
| mean | (x+y)/2 | favour class with highest confidence biggest value for decision | 24.89% | 26.60% |
| symmetrical sums | $\sigma_0$ | associative | 23.54% | 25.31% |

where N represents the number of queries and $q_i$ represents the average precision score for the $i-th$ query. Its value gives us the relevance of the retrieved documents.

## 4. Clustering

Due to the size of real medical image databases in a hospital, the performance of the retrieval process is far from real time. In our experiments, even for only 50,000 images and despite using advanced data structures, the retrieval time is about 30 seconds per query. The challenge is then to work with much more important datasets (like those existing actually in the main hospitals) reaching response-times close to real-time (at least less than 30 seconds, the classical Google failure threshold).

An interesting way to achieve this goal is to use clustering methods at the semantic level to create meaningful data-subsets. This method is able to decrease the search time by avoiding the scanning of the whole database for each query. Another interesting added value of the clustering in our study, is the semantic clustering, i.e., the possibility to have a latent meaning associated to the clusters generated.

One of the most well-known clustering technique is the k-means. The instability of ordinary k-means technique, linked to the choice of the k cluster centers that are randomly initialized, can be solved by the Fuzzy Min-Max (FMM) technique [16~18].

In this study, we explore the use of this FMM technique to determine and initialize the k-centers in an iterative way. During this initialization phase, n-dimensional hyper-cubes are created. These hyper-cubes are limited by the maximal and minimal coordinates of each hyper-cube points. A membership degree of a point for each hyper-cube is defined as follows:

$$H_j(x, v_j, u_j) = \frac{1}{n}\sum_{i=1}^{n}[1 - f(x_i - u_{ji}) - f(v_{ji} - x_i)]. where f(x) = \begin{cases} 1, & x > \eta \\ x/\eta, & if\ 0 < x \le \eta \\ 0, & x \le 0 \end{cases} \quad (7)$$

with:
$H_j$ the membership degree of the point x to the hyper-cube j. This membership degree belongs to [0,1];
$x_i$ the dimension of the input vector x;
$u_{ji}$, $v_{ji}$ the maximal and minimal value of the $i^{th}$ dimension for the $j^{th}$ hyper-cube.

The $\eta$ parameter is called the sensitivity of each hyper-cube. This parameter determines the decreasing of the membership degree $H_j$ (Equation 7) of a point according to its distance from the $j^{th}$ hyper-cube. The authors of this algorithm do not give a formal way to determine the sensitivity $\eta$. The only criterion is to minimize the overlap between hyper-cubes (small values of $\eta$). To respect this criterion we propose to calculate the value of the parameter $\eta$ using:

$$\eta = \min_{i} \frac{\max_{x_j \in \chi}(x_j)_i - \min_{x_j \in \chi}(x_j)_i}{2 \times (N-1)} \quad (8)$$

with:
$x_{j_i}$ the dimension of the input vector $x_j$ of the cluster $\chi$;
N the number of points of the cluster $\chi$.

The Fuzzy Min-Max algorithm has three phases: the expansion of hyper-cubes, overlap test and contraction of hyper-cubes:

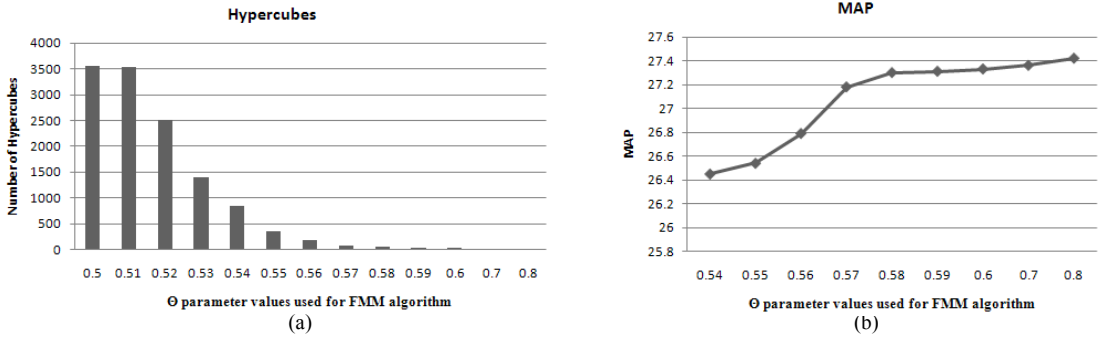

**Fig. 5** Study on the influence of the $\theta$ parameter on FMM algorithm for PEIR (CLEF) database. (a) Number of hypercubes depending on the $\theta$ parameter. (b) MAP on the $\theta$ parameter.

- initialization of the maximum and minimum points of the first hyper-cube with the first presented point;
- computation of the membership degree of each input point by Equation 7;
- expansion of the hyper-cube having the highest membership degree is done according to:

$$\sum_{i=1}^{n}(\max(u_{ji},x_i)-\min(v_{ji},x_i))\leq n\theta \quad (9)$$

Where $\theta$ represents a parameter of the algorithm which controls the creation of new hyper-cubes.

- if no hyper-cube can be expanded (condition of Equation 9 is not satisfied), a new hyper-cube containing the new input data will be generated.

In order to find the optimal value for the parameter $\theta$, we test (**Fig. 5**(a)) several configurations with respect to reasonable clustering time. This parameter is very important as it influences indirectly the retrieval values as shown in **Fig. 5**(b). Tuning this parameter is essential as it balances the clusters. The precision of the final clusters influences the retrieval result. Thus, the parameter $\theta$ represents the equilibrium between retrieval precision/recall and computation time. Our tests have revealed an achievement of 20 seconds/query, on CLEF database. This reduction in query processing time obviously comes with a compromise in recall/precision performance.

The clustering of the fused database is done off-line, inducing an increase in the on-line retrieval performance. We compute the similarity only on the relevant clusters (H>0).

## 5. Retrieval

The fused data is used in the retrieval process. The most frequently used similarity function is the KNN (K Nearest Neighbor) [19]. The similarity degree can be computed in various manners.

In [20], we summarize the similarity functions most relevant to our study. The Cosine, Jaccard and Dice similarity functions are often used for text documents since they are based on tf/idf. After fusion, the meaning of indexes has evolved. It is interesting to use the co-occurrence since it also includes the distance factor. Note that the VSM function applies to pseudo-documents. In our case, the fused image and text information could be considered as a kind of pseudo-document, but the vector must contain all the possible combinations of items. We consider also consider the distance-based similarities, used for image similarity, where the Quadratic form and the Mahalanobis distance take into account the similar pairs of bins.

We introduce a new similarity function (**Table 2**), called Fuzzy Similarity Function (FSF), a mixed similarity function based on the fuzzy operators *max* and *+*. The global confidence degree obtained by applying this similarity function takes into account the union of the most optimistic confidence degree for image and text common concepts.

$$sim_{FSF}(\lambda_q,\lambda_d)=\sum_{t\in q\cap d} A\cdot max(\lambda_{q,t},\lambda_{d,t}) \quad (10)$$

Table 2 Similarities measures used.

| Similarity measure | MAP(CLEF) |
|---|---|
| Cosine - used for document retrieval | 18.53% |
| Dice - used for document retrieval | 17.17% |
| VSM - used for pseudo-documents | 15.59% |
| FSF - takes into account the similar items | 25.32% |

where A is the similarity matrix of feature vectors, being the result of the $\cap$ operator on the CUIs extracted from the text and the image files (for the common CUIs from the query and the medical case the value in the matrix is 1, for the rest is 0). The $\lambda$ values are computed using equation 1. The $\lambda_q$ and $\lambda_d$ are respectively the $\lambda$ values for the query and the compared document from the database.

In our study, we apply a selection of similarities measures represented in **Table 2** in the final step of our fusion for searching in the database for the similar documents, depending on the concepts extracted from the database. The similarity formulas from this table have been applied for the fused values from the concepts from a document inside the database and the query document. It must be stated that a document in this case represents the concepts from the text and the image that have been fused and are stored as dictionaries. The aim of these formulas is to find the documents that have the most concepts in common with the query document, giving a higher similarity degree such that they are retrieved in a correct order.

### 6. Results on CLEF medical image retrieval benchmark

Our approach is developed using the CLEF[*10] medical image collection and tested with the Cross Language Image Retrieval track. This database contains four public datasets (CASImage, MIR, PathoPic, PEIR[*11]) that contain 50,000 medical images with associated medical reports in three different languages. Some of the tests are deployed also on PEIR database only (33,000 images), in order to enhance our contribution on homogeneous databases.

In the pre-processing phase we conducted a comparative study on the $\omega$ (spatial localization fuzzy weight) and $\varphi$ (data test feedback or relevance feedback) parameters, using small variations around a theoretically suitable structure, composed by the *sum* fusion operator (simple, commutative, associative and balanced technique) and the Fuzzy Similarity Function (FSF) for the similarity.

We compute the Mean Average Precision (MAP) of the retrieval on PEIR database from CLEF 2006 medical image retrieval benchmark, using different ($rp$) values from 0.00 to 0.60 (i.e, precisions corresponding to 60% of relevant cases retrieved) and we obtained the results presented in **Fig. 6**. The best alignment configuration corresponds to the precision associated to 30% of the relevant cases retrieved ($rp$ = 0.30). This experimental result indicates that 30% of the component media constitutes a pertinent subspace to design an optimum fusion configuration. The results using the probabilistic and fuzzy fusion techniques are presented in the last column of **Table 1**. The evidence-based techniques were not used because they are more appropriate for the decision-based fusion. From our experiment, we can deduce that the T-norms performs the best on CLEF 2006 queries/database.

We compare classical retrieval functions (Cosine, Dice, VSM) with the FSF. The obtained results are presented in the last column of **Table 2**, revealing the effectiveness of the FSF. Comparative results with other mixed automatic CBMIR approaches recorded in CLEF 2006 medical image retrieval track are given in **Table 3**. Without using any filtering (which is the case of the first one), our method has the second best performance, where the first uses a mixed approach with a multiple filtering technique. Considering that the tests on the automatic text retrieval are around 22,55% in MAP and the automatic image retrieval around 6,41% in MAP for the same indexes, the fusion applied here is effective since it gives a result greater than the sum of image and text partial retrieval results. The main screen-shots of the fusion prototype built for this purpose is represented in the **Fig. 7**.

---

[*10] CLEF - Cross Language Evaluation Forum – www.clef-campaign.org/
[*11] www.casimage.com, gamma.wustl.edu/home.html, alf3.urz.unibas.ch/pathopic/intro.html, peir.path.uab.edu

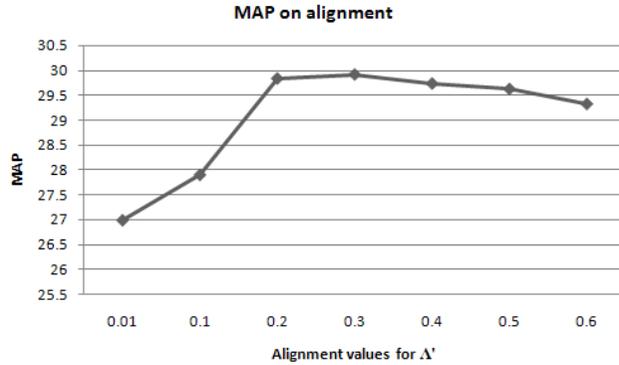

**Fig. 6** MAP for different partial Interpolated recall - precision averages. Study for PEIR database.

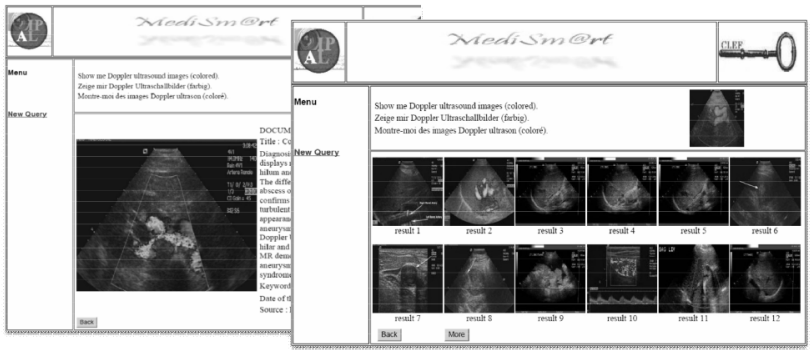

**Fig. 7** Interface of a fusion CBMIR prototype illustrating the query top 12 responses with the focus on the medical image.

**Table 3** Results on CLEF 2006.

| Run | MAP [%] | R-prec [%] |
|---|---|---|
| IPAL-IPAL_Cpt_Im.eval | 30.95 | 34.59 |
| Fusion on PEIR database | 29.91 | 31.68 |
| Fusion on CLEF database | 25.59 | 29.65 |
| UB-UBmedVT2.eval | 20.27 | 22.25 |
| RWTHi6-EnFrGePatches | 16.96 | 20.78 |

### 7. Conclusion

　Dubois and Prade[21] showed that multimedia data features are often characterized by the lack of standard structure, heterogeneity of formats, self describing information, inadequacy of textual descriptions, multiplicity of data types and source database, spatial and temporal characteristics. Consequently, all fusions methods still remain strongly related to the accuracy and the structure of the initial input data.

　Nevertheless, the richness of the UMLS metathesaurus and the results on CLEF medical image retrieval benchmark open interesting perspectives for future developments related to semantic clustering and query expansion to improve the robustness, efficiency and effectiveness of medical image retrieval systems.

　Our first results in the clustering of the fused indexes represent an interesting idea for the future, as it reveals opportunities to improve the computation time/accuracy rate. The clustering applied in the semantic frame also has potential applications to medical multimedia data mining.

　The global performances are obviously improved for an homogeneous database (i.e. PEIR versus CLEF - see **Table 3**). A similar conclusion can be extended for the clustering results (MAP 25.59% after clustering on CLEF and MAP 27.23% after clustering on PEIR). This highlights again the importance of the initial input data structure; the more specialized and homogeneous the database, the more efficient the retrieval.

The reduction of the on-line and off-line computation time, especially for the medical image indexing, may also use the grid computing facilities. Such a grid-enabled medical image retrieval system will be studied in the frame of the ONCO-MEDIA project to improve the performances of CBMIR systems.

**Acknowledgment**

This work has been done with the support of ONCO-MEDIA[*12] ICT Asia project. We would like also to thank our colleagues from IPAL - Caroline Lacoste, Nicolas Vuillemenot, Le Thi Hoang Diem and Jean-Pierre Chevallet - for providing us the text and images separate indexes used for the experimental part. For the financial support for the publication we would like to thank the Kayamori Foundation of Informational Science Advancement.

\*    \*    \*